\documentstyle[12pt]{article}

\begin{document}
\begin{sf}
\begin{center}
 DOMAIN PATTERNS IN INCOMMENSURATE SYSTEMS WITH THE UNIAXIAL REAL ORDER
 PARAMETER\\
 V. Danani\'{c}\\
 Department of Physics \\
 Faculty of Chemical Engineering and Technology, \\
 University of Zagreb, \\
 Maruli\'{c}ev trg 19, 41000 Zagreb,  Croatia\\
 and \\
 A. Bjeli\v{s}\\
 Department of Theoretical Physics, Faculty of Sciences,\\
 University of Zagreb, \\
 POB 162, 41001 Zagreb, Croatia\\

\end{center}
 PACS numbers: 64.70.Rh ; 64.60.My
\newpage
\begin{center}
 Abstract
\end{center}
 The basic Landau model for the incommensurate-commensurate transition to the
 uniform or dimerized uniaxial ordering is critically reexamined. The previous
 analyses identified only sinusoidal and homogeneous solutions as
thermodynamically
 stable and proposed a simple phase diagram with the first order phase
transition
 between these configurations. By performing the numerical analysis of the free
energy
 and the Euler-Lagrange equation we show that the phase diagram is more
complex. It also
 contains a set of metastable solutions present in the range of coexistence of
homogeneous
 and sinusoidal solutions. These new configurations are periodic patterns of
homogeneous
 domains connected by sinusoidal segments. They are Lyapunov unstable, very
probably due
 to the nonintegrability of the free energy functional. We also discuss some
other
 mathematical aspects of the model, and compare it with the essentially simpler
sine-Gordon
 model for the transitions to the states with higher commensurabilities. The
present results
 might be a basis for the explanation of phenomena like thermal hystereses,
cascades of phase
 transitions, and memory effects, observed in materials exhibiting the
transitions to uniform
 or dimerized state. Some particular examples are discussed in detail.

\newpage
\section{Introduction}
\label{sec-intro}

 During the past fifteen years there have been
 discovered over a hundred materials [1-3]
 exhibiting incommensurate properties. Up to present days there is however
 no unique microscopic  theory, and consequently one cannot give an
 universal definition of incommensurate phases. Provisionally an
 incommensurate phase can be defined as that having some
 measurable property which is periodic in space with a
 period which cannot be recognized
 as a simple (rational) multiple of underlying lattice
 periodicities. It is generally accepted that the reason for the
 occurrence of incommensurate
 phases lies in the existence of two or more competitive
 interactions which prefer different periods of ordering.
 Still, these periods may be, and in many cases
 they are, commensurate among themselves and with lattice spacings.
 The experiences on a number of models show that, depending on the relative
 strengths of interactions, the ordering may be incommensurate,
 commensurate or chaotic.

 A simple example is an array of atoms with a lattice constant $a_{0}$,
 connected with harmonic springs and interacting with a periodic background
 potential with a period $b$, $a_{0}$ and $b$ being generally incommensurate.
 This is a model proposed by Frenkel and Kontorowa (FK) \cite{kn:Frenkel}
 and studied so far by many authors \cite{kn:Bak}.
 Depending on the relative strength of the harmonic and background potentials,
 the atoms will arrange in a modulated periodic configuration with a period
 $a$ incommensurate with both $a_{0}$ and $b$, or to a commensurate
 configuration, such that $a$ is a simple rational fraction of $b$. There is
 also a third possibility which involves some mathematics developed in last
 thirty years. It concerns the chaotic behavior of the deterministic systems
 among which the FK system is one of the most intensively studied. Namely, if
 the background potential is large enough with respect to the harmonic one,
 the atoms will arrange in a chaotic way. It is important to stress here that
 the naive continuum versions of the FK model lead in the lowest order to
 the sine-Gordon equation which is $always$ integrable. There is no room
 left for chaotic solutions. So, we shall keep in mind that by the
 continuation of the original discrete space an important, chaotic feature
 of the system may be lost.

 On the other hand, the space continuation
 is a necessary step in the Landau expansion of the
 free energy. The latter is based on two approximations; the reduction of the
 summation in the reciprocal space to the narrow range around a given star
 of wave vectors, and the truncation in the series of powers in order
 parameter. While the choice of the star is related to the details of the
 competitive interactions, the introduction of the cut-offs in the wave
 vector summations is justified by the weakness of these interactions.
 Indeed, only then one can exclude distances in the real space shorter
 than reciprocal cut-offs. The only short-range scales which then remain
 are periods of fast
 modulations of order parameters, defined by the star of wave vectors
 itself.

 The truncation of power series in the of Landau expansion is strictly
 justified only in the vicinity of the phase transition
 at the temperature $T_{I}$ from
 the high temperature disordered state to the ordered one. Still,
 the Landau expansion sometimes gives a reasonably good description
 of the incommensurate-commensurate
 transitions also well below $T_{I}$. The range of its applicability
 depends on its microscopic origin. E.g. for the anisotropic
 Ising model with nearest and next nearest neighbors (ANNNI) the Landau
 expansion which follows after continuing both the space and the Ising
 variables, reproduces well only the transition line $T_{L}$
\cite{kn:Bak,kn:Selke}. On the other
 hand, the Landau expansions for electronically driven charge and spin density
 wave [C(S)DW] instabilities are as well useful in the range of lower
 temperatures below $T_{I}$. The well-known example is the rich phase
 diagram of the organic chain conductor TTF-TCNQ \cite{kn:Barisic}.

 In the Landau model the commensurate state is
 characterized by the critical temperature of IC transition
 $T_{L}$ and the three rational numbers $(r_{1},r_{2},r_{3})$ defining
 the wave vector $\vec{q_{c}}$ and the corresponding star below $T_{L}$.
 Due to the translational symmetry ({\em i.e.} periodicity) of the crystal,
 these rational numbers are placed in the interval $- \frac{1}{2} \leq r_i \leq
\frac{1}{2}$.
 The values $r_{1}=r_{2}=r_{3}=0$ and $r_{i}=\pm \frac{1}{2}$ ($i=1,2,or 3$)
 represent the center and the border of Brillouin zone respectively.
 The commensurate order below $T_{L}$ is characterized by the order parameter
 with the number of components proportional to
 the number of arms in the star of wavevectors generated by the
 space group acting on ${\vec q}_c$ \cite{kn:Landau}. In most cases the order
parameter
 is a multicomponent ($e.g.$ complex) quantity. It may be a real ($i.e.$ one
component)
 quantity only when ${\vec q}_{c}$ lies in the center or at the border of
 Brillouin zone.
 The simplest, but still very frequent, cases which allow both above
 possibilities, are those with an
 uniaxial modulation characterized by a scalar physical quantity,
 like the local electron density in charge density waves,
 the atomic displacement along some fixed direction in ferroelectric
 systems, the magnetization along some preferred easy axis in magnetic systems,
etc.
 In the present work we consider possible phases of such uniaxial ordering
 by starting from the Landau models for the free energy
 and investigating the thermodynamically stable
 configurations among the solutions
 of the corresponding Euler-Lagrange (EL) equations.

 With a finite value of ${\vec q}_{0}$ one usually associates the order
 parameter to an irreducible representation defined by the corresponding
 star of wave vectors [e.g ref.\cite{kn:Landau}]. Thus for an uniaxial ordering
 the star has two points ($\pm q_{0}$)as far as $q \neq 0,\frac{1}{2}$.
 The order parameter
 $\Psi$ is then either complex (as is most often a case), or has an even number
 of components greater than two (e.g. six for the uniaxial spin density wave
 \cite{kn:Bjelis}).
 In the former case $\Psi =\rho e^{i\Phi}$, with $\rho$ and $\Phi$ being
 the amplitude and phase which generally may vary in space already in the
 stable equilibrium state. The invariants which enter into the Landau expansion
 are the isotropic (phase indenpendent) even powers of the absolute value of
 $\Psi$ [with the coefficient in front of the quadratic term measuring e.g. the
 height of the minimum in the dispersion curve of the soft branch of lattice
oscillations,
 $\omega(q)$ (see e.g. the case of $K_{2}SeO_{4}$ \cite{kn:Haque}) at
$T>T_{I}$], and the squared
 absolute value of the gradient $\partial \Psi/\partial x$ [with the lowest
 order term $c{\mid \partial \Psi/\partial x \mid}^{2}$, where $c$ measures
e.g.
 the parabolic width of the minimum of $\omega(q)$ at $ q_{0}$]. Were the order
 parameter introduced via some other star with wave numbers
 close to $\pm q_{0}$, one would
 also have the Lifshitz invariant $\Psi \partial {\Psi}^{*}/\partial x -
 {\Psi}^{*}\partial \Psi/\partial x$ in the Landau expansion. It would
 simply account for the shift of the wave vector with respect to the true
 minimum of free energy at $ q_{0}$.

 The anisotropic, $i.e.$ phase dependent, terms appear in the Landau expansion
 when $q_{0}$ is close to the commensurate value $q_{c}=2\pi/n$
 ($n=3,4,..$). Then the original discrete dependence of the interaction
potential
 on the lattice site generates an Umklapp term which after the continuation
 of the space variable has the form
\begin{equation}
 \rho^{n} \left[e^{i[n\varphi+n(q_{0}-2\pi/n)x]}\right]
\label{eq:Umklapp}
\end{equation}
 As was already pointed out, the continuation procedure, and the approximation
 by which only one Umklapp contribution is singled out, are appropriate only
 for systems with weak local interactions. The explicit dependence of the
 term (\ref{eq:Umklapp}) on $x$ is slow, since the difference $q_{0}-2\pi/n$ is
 presumably small. This slowness is the decisive reason for retaining of
 the term (\ref{eq:Umklapp})
 in the Landau expansion. If one now concentrates on the portion of the
 phase diagram with the IC transition to $q=q_{c}$,
 it is most convenient to pass to the expansion with respect to the star
 ($\pm q_{c}$). With this choice one has the term
\begin{equation}
  \rho^{n} \left(e^{in\Phi}+ c.c.\right)
\label{eq:star}
\end{equation}
 instead of eq.(\ref{eq:Umklapp}), and the Lifshitz term with the prefactor
 $\sim(q_{0}-2\pi/n)$ in addition. Note that $\phi$ and
 $\Phi\equiv\phi+(q_{0}-q_{c})x$ are phases
 with respect to $q_{0}$ and $q_{c}$ respectively. Thus, the
 inclusion of the Umklapp term into the Landau free energy generates the
 Lifshitz invariant which is essential for the thermodynamics of the
corresponding
 IC transition. Taking into account only phase variations in space one obtains
 a well-known sine-Gordon equation as the Euler-Lagrange equation for the
 stable configurations. The corresponding phase diagram shows a "weak"
 singularity at the continuous transition from the dilute lattice of solitons
 in $\Phi$ to the commensurate  $(\Phi=cte)$ configuration
\cite{kn:McMillan,kn:Bulaevskii}.

 The approach via the above sine-Gordon model exhausts all IC transitions
 with $n>2$ in the systems with weak interactions. Of course, the cases
 $n=3$ and $n=4$ are the most interesting, since the corresponding prefactors
 of $e^{in\Phi}$ terms contain powers of amplitude compatible with the
 isotropic $\rho^{4}$-term which is unavoidable in any Landau expansion.
 For the same reason the terms coming from higher commensurabilities are
 less and less important as $n$ increases. This is in contrast to the
 strong-coupling systems in which the discrete space dependence a priori
 includes all commensurabilities.

 Two cases which are not covered by the sine-Gordon model are the homogeneous
 ordering with the periodicity of the underlying lattice $(n=1,\,\,i.e\,\,
q_{c}=0)$
 and the dimerization ($n=2$, $i.e.$ $q_{c}=\pi=-\pi(mod2\pi)$).
 In both cases the commensurate star has only
 one leg, so that the basic irreducible representation is one-dimensional,
 defining a real order parameter. The notion of phase then looses its sense.
 This may be expressed in the following more descriptive way.
 For all periodicities
 (incommensurate, commensurate with $n>2$ and commensurate with $n=1,2$) one
 may characterize the sinusoidal modulation by its amplitude and by the
relative
 phase with respect to the lattice. However, in contrast to other
periodicities,
 in the case of commensurate orderings with $n=1,2$ this phase
 just determines the amplitude of the modulation, which is same (for $n=1$)
 or has same absolute value (for $n=2$) at all lattice sites. In other words,
 the relative phase is "absorbed into the amplitude", $i.e.$ it is already
 completely defined by the value of the modulation at the lattice site.
 This value remains the only quantity which characterizes the ordered
 phase, representing the real order parameter $u$.
 Since, as is indicated above, for $n=1,2$ all lattice sites are equivalent,
 it is possible to formulate a common generic Landau expansion for both cases.
 As will be seen in Sec.2, the essential feature of this expansion is the
 possible negative sign of the term proportional to the squared first
 derivative of the order parameter. In order to stabilize the expansion
 one then has to include the term with the squared second derivative. This
 completes a minimal model for the commensurate lock-ins with $n=1,2$. It can
be extended
 by further possible invariants, depending on the details in particular
 physical systems. The present analysis will be however limited to the
 minimal model. Its main aim is to point out basic qualitative differences
 between IC transitions with $n=1,2$ and those with $n>2$.

 The division of uniaxial IC transitions into two classes $i.e.$ those with
 (class I) and without (class II) Lifshitz invariant is well established
 in the literature  \cite{kn:Blinc,kn:Cummins,kn:Bruce1,kn:Toledano}. While the
 sine-Gordon model, as a minimal one for the former class, can be completely
 integrated \cite{kn:McMillan,kn:Bulaevskii}, the problem of integrability
 of the model for the latter class is far from being resolved. Mathematically,
 the models based on the Lagrangian with higher derivative terms are
 equivalent to the Hamilton problems with unbounded kinetic part. Thus one
 misses a visualization of corresponding solutions through some mechanical
 analogues. In spite of these difficulties, the phase diagram for the class II
 suggested in previous works (see e.g. \cite{kn:Bruce1,kn:Michelson})
 is surprisingly simple, even simpler than that of
 sine-Gordon model. It includes only two homogeneous solutions, $u=0$ and
 $u=u_{0}$, representing the disordered and commensurately ordered
 configurations respectively, and an (almost) sinusoidal solution $u(x)$
 with an incommensurate wave number. The IC transition itself is of the
 first order. As it was explicitly stated \cite{kn:Bruce1},
 the most significant property of this phase diagram
 is absence of configurations of
 soliton-lattice type, in contrast to the corresponding phase diagram
 of the sine-Gordon model.

 In the present work we start from the question of the possible presence
 of patterns consisting of alternating commensurate domains, $i.e.$ of
 soliton lattices for the class II systems. Due to the afore mentioned
 nontrivial mathematical properties of EL equation, such periodic solutions
 might cover only tiny subsets in the complete space of solutions.
 Under these circumstances a rather careful numerical analysis is
 unavoidable. Combining two indenpendent numerical methods, we find a new
 sequence of highly nonsinusoidal solutions having a form of domain patterns.
 They are locally stable ($i.e.$ metastable) in the range of control
 parameter in which the commensurate configuration is metastable
 ("superheated"). As will be argued below, the presence
 of these solutions substantially enriches the phase diagram and may have
 direct consequences on the IC transitions in real systems.

 In Sec.2 we formulate the Landau model for the II class and discuss
 its mathematical properties. In Sec. 3 we introduce the systematization
 of periodic solutions and describe the numerical procedures for obtaining
them.
 The structure of the phase diagram is presented in Sec.4. Finally, in the
concluding Sec.5 we
 point out some wider theoretical aspects of our results and discuss possible
implications on
 experimental properties of few well-known materials from the II class.
\newpage

\section{The model and its mathematical aspects}
\label{sec-model}
 The wave number for the uniaxial system with a real order parameter $u(x)$
 is situated either at the center ($q=0$, for $n=1$) or at the border
 ($q=\pi/a$ for $n=2$) of the first Brillouin zone. The simple transformation
 by which in the latter case the origin of the wave number is shifted by
 $\pi/a$, enables the construction of common Landau expansion for both cases.
 This transformation corresponds to the elimination of fast variations in the
 direct space, realized by passing from the "displacement" at the $n-$th site,
 $d_{n}$, to $u_{n}=(-1)^{n}d_{n}$. Since $u_{n}$ slowly varies on the lattice
 scale, just like the "displacements" in the $n=1$ case, one may pass to the
 continuous space coordinate providing that the interactions are weak
 enough.

 The essential novel property of the Landau expansion for the incommensurate
 order close to $n=1$ or $n=2$ follows from the dependence of the free energy
 density on the wave number $q$. The quadratic part $f_{2}(q)$ has to be an
 even function of $q$. Thus, as long as local minima of $f_{2}(q)$ are at
 finite values $\pm q_{0}$, it has a shape of bottle bottom, modelled by
\begin{equation}
 f_{2}(q)=\left(a+cq^{2}+{\scriptstyle \frac{1}{2}}dq^{4}\right)u^{2}(q)
\label{eq:bottle}
\end{equation}
 with $c<0$, and $q_{0}^{2}=-c/d$. Here $d>0$ by assumption. The corresponding
 expression for the free energy functional in the direct space is
\begin{equation}
 f[u]=\frac{1}{2L}\int_{-L}^{L}\left[d\left(\frac{d^{2}u}{dx^{2}}\right)^{2} +
 c\left(\frac{du}{dx}\right)^{2} + au^{2} + {\scriptstyle \frac{1}{2}}
bu^{4}\right]dx
\label{eq:fe1}
\end{equation}
 where $L$ is some macroscopic length ("volume"). Here $b>0$, so that the
 first and the fourth term in eq.(\ref{eq:fe1}) ensure that the free energy
 functional is bounded from below. The coefficients $a$ and $c$ depend on
 temperature and perhaps on some other physical parameter(s). It is expected
that there
 is a physical regime in which both coefficients are negative. The model
 (\ref{eq:fe1}) may be also formulated in the frame of
 the Lifshitz point theory \cite{kn:Hornreich,kn:Michelson}.
 It is a minimal one since no higher order derivative terms like
 $(d^{3}u/du^{3})^2$, $u^{2}(du/dx)^{2}$ etc. are included.
 Some authors \cite{kn:Yoshihiro} studied
 a more extensive model in which
 the invariant of the form $u^{2}(du/dx)^{2}$ is taken into account.
 The extensions of this kind may enrich the phase diagram, simply
 because the dimension of control parameter space is larger.
 However, they do not alter qualitatively the
 main conclusions of our analysis.

 Some conclusions about stable configurations $u(x)$ can be gained from
 a direct consideration of the free energy functional (\ref{eq:fe1}).
 Up to "surface" terms, it
 can be recast into the form
\begin{equation}

f[u]=\frac{1}{2L}\int_{-L}^{L}\left[d\left(\frac{d^{2}u}{dx^2}-\frac{c}{2d}u\right)^{2}+
 \left(a-\frac{c^2}{4d}\right)u^2+{\scriptstyle \frac{1}{2}} bu^{4}\right]dx
\label{eq:fe2}
\end{equation}
 which is convenient for the following reasoning.
 In reaching thermodynamical equilibrium, the first term in eq. (\ref{eq:fe2})
 prefers sinusoidal modulation of $u(x)$ with the wave vector
 $q=\sqrt{-c/2d}$, provided that $c<0$. The remaining two terms would then
stabilize
 the amplitude of the oscilations as long as $a-c^{2}/4d<0$.
 This simple reasoning gives us an idea how and for what values of control
 parameters $a$ and $c$
 imcommensurate phases could emerge.
 Indeed, the line $a=c^{2}/4d$ in the phase diagram
 for the free energy (\ref{eq:fe1}) represents
 the transition line between the disordered phase $u(x)=0$
 and the incommensurate sinusoidal
 phase with the wave vector $q=\sqrt{-c/2d}$. This conclusion is exact at the
 transition line, but becomes approximate for $a<c^{2}/4d$ since the
 modulation does not remain purely sinusoidal, so that the higher harmonic
 terms have to be included in order to reach a true minimum. However, such
 terms represent only a small correction to both the free energy and the
 wave vector \cite{kn:Michelson} (see also Sec.4 below).
 On the other hand, further, more complex periodic configurations which
 might minimize the free energy functional well below the line
 $a=c^{2}/4d$ cannot be found as a mere extension
 of this almost sinusoidal configurations.
 In this respect we note that there is a freedom in the rearrangement of terms
 of the functional (\ref{eq:fe2}) by introducing various "surface"
contributions.
 Although it cannot affect thermodynamic properties
 of the functional(like $e.g.$ the existence of stable and metastable
 configurations), this freedom of choice may suggest other
 periodic solutions as candidates for (meta)stable configurations, besides
 that which follows from the choice (\ref{eq:fe2}). Evenmore, one can pose a
 question whether in such a way the non-periodic ($i.e.$ multiperiodic and
 even chaotic) configurations can emerge as possible configurations which
 at least approximately minimize the free energy \cite{kn:surface}.
 In any case, the above type of reasoning is helpful when corresponding
 EL equation cannot be solved analytically, since it sometimes gives
 plausible guesses for the initial dependence of $u(x)$ and so facilitates the
 numerical procedure in the calculation of true solutions.

 We proceed by considering some mathematical aspects of the problem.
 Since the coefficients $b$ and $d$ in the equation (\ref{eq:fe1}) are positive
by
 assumption, we can redefine the order parameter and free energy density by
 introducing
\begin{eqnarray}
 u(x)&=&\sqrt{\frac{d}{b}}\,\overline{u}(x), \nonumber \\
    f&=&\frac{d^2}{b}\,\overline{f},         \nonumber \\
    c&=&d\,\overline{c}{\;,\;}a=d\,\overline{a},
\label{eqnarray:coef}
\end{eqnarray}
 so that
\begin{equation}
 \overline{f}[\overline{u}]=\frac{1}{L}\int_{0}^{L}\left[{\left(\frac{d^{2}
 \overline{u}}{dx^{2}}\right)}^{2}+\overline{c}{\left(\frac{d\overline{u}}{dx}

\right)}^{2}+\overline{a}\,\overline{u}^{2}+{\scriptstyle{\frac{1}{2}}}\overline{u}^{4}\right]dx.
\label{eq:feren}
\end{equation}
 In what follows we shall omit the bar above the quantities appearing in
 eq.(\ref{eq:feren}).

 The search for thermodynamically stable configurations $u(x)$
 begins with the study of the EL equation for the functional
 (\ref{eq:feren})
 as the necessary condition which each such configuration
 has to obey. It reads
\begin{equation}
 \frac{d^{4}u}{dx^{4}}-c\,\frac{d^{2}u}{dx^{2}}+a\,u+u^{3}=0.
\label{eq:EL}
\end{equation}
 The acceptable solutions $u(x)$ of this equation are those for which the
 corresponding eigenvalue equation
\begin{equation}

\frac{d^4\eta(x)}{dx^4}-c\frac{d^2\eta(x)}{dx^2}+(a+3u^{2}(x))\eta(x)=\lambda\eta(x)
\label{eq:eig}
\end{equation}
 with boundary conditions
\begin{eqnarray}
 \eta(0)=\eta(L)=0 \nonumber \\
 \eta^{\prime}(0)=\eta^{\prime}(L)=0
\label{eqnarray:eigbc}
\end{eqnarray}
 generates only positive values of $\lambda$'s. This is
 the sufficient and necessary condition for the thermodynamic stability of
 the solution $u(x)$.

 We are not aware of any method which would lead to a complete integration
 of equation (\ref{eq:EL}).
 A straightforward analysis of equation shows that it does not
 possess the Painlev\'{e} property $i.e.$ its movable singularities are not
 only simple poles in the complex x-plane. The Laurent expansion of the
 solution which starts with a simple pole does not exist, so that logarithmic
 terms have to be included.
 In such cases one may try a general transformation of both the
 dependent ($u$) and independent ($x$) variables with
 the aim to find the transformed differential
 equation which reveals Painlev\'{e} property.
 We have not found any such
 transformation. Even if it exists, it would
 not prove the integrability of (\ref{eq:EL}) but would give only a strong
 indication for it. So, from the point of view of Painlev\'{e} analysis we
cannot
 conclude anything about integrability of equation (\ref{eq:EL}) except that it
leaves
 us with strong impression of its probable nonintegrability.

 Alternatively one could look for a sufficient number of functionally
 independent invariants ("integrals
 of motion") of differential equation (\ref{eq:EL}) in
 order to reduce it to a differential
 equation of first order, integrable by quadrature. Since
 we deal with an autonomous differential equation derived from the variation
 of a functional, there is certainly one such invariant. It has the form
\begin{equation}
 H=\frac{1}{2}u^{4}+au^{2}-cu'^{2}-u''^{2}+2u'u'''.
\label{eq:H}
\end{equation}
 The function $H$ is analytic in all four variables. It is then
 natural to ask whether there exists another analytic invariant. We have
 tried to find it by performing Taylor expansion in four
 variables. Such formal Taylor expansion
 exists only in cases when all roots of the characteristic
 equation of the linear part of (\ref{eq:EL}) are different. By $formal$ one
means
 that its coefficients contain inverse powers of the integer combinations
 of the four roots. If the Taylor series is infinite it cannot be convergent
 since among these integer combinations there are always those which are
 vanishingly small (or are even equal to zero if the roots of
 eq.(\ref{eq:EL}) are commensurate). The only possibility to overcome such
difficulties
 is to require the termination of Taylor expansion $i.e.$ to search for
 other polynomial invariant. It can be however shown that $H$ given by
 eq.(\ref{eq:H}) is the only possible polynomial invariant.
 In cases when the roots are degenerate there
 exists no other analytic invariant at all. At the time being, we cannot
 say anything about nonanalytic invariants.

 The equation (\ref{eq:EL})
 can be put in the form of Hamilton system of equations with two degrees
 of freedom. $H$ from equation (\ref{eq:H}) is then the value of the
 Hamiltonian for the given solution $u(x)$.
 This property ensures the sufficiency of two invariants
 for the integrability. Since we cannot either find or prove
 the existence of the second invariant,
 there is no definite answer to the question of
 integrability of equation (\ref{eq:EL}).

 Passing to the numerical treatment of eq.(\ref{eq:EL}), we note at the
 beginning that an $ad$ $hoc$ choice of initial conditions as a rule leads
 to the unbounded solution $u(x)$. The solutions
 which grow beyond any limit cannot have a physical meaning, because they
 do not obey the requirements for the validity of Landau expansion.
 Let us choose, for example, $c=-1$ and $a=-1$. There are two thermodynamically
 stable solutions and one unstable homogeneous solution,
 $u(x)=\pm1$ and $u(x)=0$ respectively.
 The determinantal equation obtained from the equation (\ref{eq:eig})
determines
 the lowest eigenvalue $\lambda$ as a function of $L$ for each solution.
 By choosing sufficiently large $L$ one recovers that for stable homogeneous
 solution these eigenvalues correspond to eigenfunctions
 $\eta(x)\propto e^{ikx}$ with real $k$. The lowest eigenvalue $\lambda$
 is then obtained by minimizing it as a function of $k$. One then tries to
 integrate the equation (\ref{eq:EL}) by using Runge-Kutta method of higher
order
 and choosing initial conditions
 very close to those for stable homogeneous solutions.
 The result is an extremely fast growing solution $u(x)$. The same result
 will be obtained for initial conditions close to the
 unstable homogeneous solution $u(x)=0$. The reason
 for such behavior can be traced by looking at the Lyapunov exponents
corresponding
 to the homogeneous solutions. If $\mu$ is one exponent
 then so is $-\mu$. The meaning of this property is that there is no Lyapunov
 stable solutions except when all exponents are purely imaginary, in which case
 the linearized equation (\ref{eq:EL}) leads to the neutral Lyapunov stability.
 In the example given above Lyapunov exponents have
 nonvanishing real parts for all homogeneous solutions.
 So, the thermodynamic stability of $u(x)$ here goes together with its
 orbital instability. This however need not be always the case.

 The counter example of a thermodynamically unstable and orbitally stable
 solution is realized for e.g. $c=-1$ and $a=-0.09$.
 The homogeneous solutions $u(x)=\pm 0.3$ are then
 thermodynamically unstable for sufficiently large $L$, but at the same
 time, the inhomogeneous solutions in their vicinity can be easily
 obtained using Runge-Kutta method. A small perturbation of initial conditions
 does not produce significant change in $u(x)$. So, the solutions sufficiently
 close to the homogeneous solutions are orbitally stable.
 Among these solutions we have not found any chaotic solution. Thus, the only
 outcome of the numerical treatment of eq. (\ref{eq:EL})
 in this case is that there are
 periodic solutions close to homogeneous solutions for which the
 corresponding Lyapunov exponents are imaginary. It is not neccesary that
 all four exponents have to be purely imaginary in order that eq. (\ref{eq:EL})
 possesses a periodic solutions. It is enough to have only one pair of
 complex conjugate purely imaginary Lyapunov exponents. However, in such
 cases the Runge-Kutta integration is difficult since the remaining two
 exponents are real and of opposite signs.

 Finally, it is instructive to calculate the free energy
 (\ref{eq:feren}) of a given solution
 $u(x)$ for various values of $L$. The function $f(L)$ obtained in this way
 tends very fast to a definite constant value although the corresponding
 configuration $u(x)$ may not represent a local minimum of free energy
 functional. Similar situation occurs around every homogeneous solution
 for which the linearized part of eq.(\ref{eq:EL}) gives purely imaginary
 Lyapunov exponents.

 For the sake of completeness, let us also designate the regions
 in the ($a,c$) parameter space in which
 particular homogeneous configurations are thermodynamically stable.
 The disordered (paramagnetic) phase $u(x)=0$ is stable for $a\geq c^{2}/4$.
 The (anti)ferromagnetic phases $u(x)=\pm\sqrt{-a}$ exist for $a<0$ and are
stable
 for $a<-c^{2}/8$.

 In cases when all Lyapunov exponents with respect to $e.g.$ given homogeneous
 solution are imaginary, one may expect to find
 two-periodic thermodynamically stable solutions in the vicinity of this
 solution. We did not find appropriate
 method to derive such solutions.
 The same holds for chaotic solutions. Thus, besides homogeneous solutions,
 there remain only periodic solutions which are numerically tractable.

\newpage
\section{Periodic solutions}
\label{sec-per}
 Any periodic solution $u(x)$ with the period $2\pi/k$
 can be expanded into a Fourier series
\begin{equation}

u(x)=a_{0}+\sqrt{2}\,\sum_{n=1}^{\infty}\left(a_{n}\cos(nkx)+b_{n}\sin(nkx)\right)
\label{eq:fs1}
\end{equation}
 This series can be further simplified after noticing that
 the differential equation (\ref{eq:EL}) is autonomous and invariant under
 reflections $x\rightarrow-x$ and $u\rightarrow-u$, so that the solutions
 may have additional symmetries. Let us distinguish four possible classes:

 class ($i$) The solutions possessing maximal symmetry in the phase plane
$(u,u')$
 $i.e.$ being even with respect to both axes, $u$ and $u'$. The corresponding
 Fourier series is given by
\begin{equation}
  u(x)=\sqrt{2}\,\sum_{n=1}^{\infty}a_{n}\cos(2n+1)kx
\label{eq:fsI}
\end{equation}

 class ($ii$) Solutions which are even with respect to the $u$-axis, but
generally
 have no symmetry with respect to the $u'$-axis. For them at least one pair
 $a_{2n},a_{2m+1}$ and/or $a_{0}$ in the expansion (\ref{eq:fs1})
 are different from zero. The Fourier series has a general form
\begin{equation}
 u(x)=a_{0}+\sqrt{2}\,\sum_{n=1}^{\infty}a_{n}\cos(nkx)
\label{eq:fsII}
\end{equation}

 class ($iii$) Solutions which are even with respect to the $u'$-axis and have
no
 symmetry with respect to the $u$-axis. For them at least one pair
$b_{2n},b_{2m+1}$
 in the expansion
\begin{equation}
 u(x)=\sqrt{2}\,\sum_{n=1}^{\infty}b_{n}\sin(nkx)
\label{eq:fsIII}
\end{equation}
 is different from zero. Finally,

 class ($iv$) contains solutions which generally have no symmetry in the
$(u,u')$
 plane. They may be represented in the following form
\begin{equation}

u(x)=a_{0}+\sqrt{2}a_{1}\cos(kx)+\sqrt{2}\,\sum_{n=2}^{\infty}\left[a_{n}cos(nkx)+b_{n}sin(nkx)
\right]
\label{eq:fsIV}
\end{equation}
 where $a_{1}$ and at least one of $b_{n}$ are different from zero.

 For any periodic configuration $u(x)$ the functional (\ref{eq:feren}) may be
 replaced by the functional
\begin{equation}

f\left[u(x)\right]=\frac{k}{2\pi}\int_{0}^{\frac{2\pi}{k}}\left[{u^{\prime\prime}}^{2}
 +c\,{u^{\prime}}^{2}+a\,u^{2}+{\scriptstyle\frac{1}{2}}u^{4}\right]dx,
\label{eq:feper1}
\end{equation}
 with an error which vanishes as $1/L$ for large $L$. Replacing further the
 integration coordinate $x$ by the coordinate
\begin{equation}
 z=kx,
\label{eq:percoor}
\end{equation}
 we come to the functional
\begin{equation}

f\left[u(z)\right]=\frac{1}{2\pi}\int_{0}^{2\pi}\left[k^{4}\,{u^{\prime\prime}}^{2}
 +ck^{2}\,{u^{\prime}}^{2}+a\,u^{2}+{\scriptstyle\frac{1}{2}}u^{4}\right]dz
\label{eq:feper2}
\end{equation}
 where $u^{\prime}\equiv du/dz$ etc.
 Finally, inserting the Fourier expansions (\ref{eq:fsI}),(\ref{eq:fsII}) and
 (\ref{eq:fsIII}) into the equation (\ref{eq:feper1}) we get
\begin{equation}
 f\left[u(x)\right]\equiv f(k,a_{0},a_{1},\ldots)=\sum_{n=0}^{\infty}\left(
 k^{4}n^{4}+c\,k^{2}n^{2}+a\right)a_{n}^{2}+\frac{1}{2}\left<u^{4}\right>,
\label{eq:feper3}
\end{equation}
 where
\begin{equation}
 \left<(\cdots)\right>\equiv\frac{1}{2\pi}\int_{0}^{2\pi}(\cdots)dz.
\label{eq:meanv}
\end{equation}
 The amplitudes $a_{0},a_{1},\ldots$ in the eq.(\ref{eq:feper3}) represent
 any of the three sets of $a_{n}$ or $b_{n}$
 amplitudes (\ref{eq:fsI}),(\ref{eq:fsII}) and (\ref{eq:fsIII}). For the
 solutions of the class ($iv$) the coefficients $a_{n}^{2}$ in the quadratic
 part of eq.(\ref{eq:feper3})  are replaced by $a_{n}^{2}+b_{n}^{2}$.

 The functional (\ref{eq:feper3}) can be immediately minimized with respect to
 the wave vector $k$ (note that $<u^{2}>$ and $<u^{4}>$ do not depend
 on $k$);
\begin{equation}
 \frac{\partial f}{\partial k}=0 \Longrightarrow 2k\left[
 \sum_{n=1}^{\infty}\left(2k^{2}n^{4}+c\,n^{2}\right)a_{n}^{2}\right]=0
\label{eq:gradfk}
\end{equation}
$i.e.$
\begin{equation}
 k^{2}\equiv k_{0}^{2}=-\frac{c}{2}\frac{\sum_{n=1}^{\infty}n^{2}a_{n}^{2}}
 {\sum_{n=1}^{\infty}n^{4}a_{n}^{2}}.
\label{eq:mink}
\end{equation}
 We note that the wave vector $k_{0}$ from the equation (\ref{eq:mink}) is
 measured in units of $\sqrt{-c}$. Furthermore, the most interesting periodic
 solutions are expected when $c<0$ and $a<-c^{2}/8$, since in this range the
 homogeneous configurations $u(x)=\pm\sqrt{-a}$ are stable. It is then
 natural to ask how homogeneous solutions participate in
 periodic solutions, $i.e.$ whether the latter have ferromagnetic
 segments. In particular,
 we expect that ferromagnetic solutions and almost sinusoidal solutions
 could be mixed into new periodic configurations.
 Let us therefore limit further analysis to the range of parameters
 $c<0,a<0$ and introduce following redefinitions:
\begin{eqnarray}
 &k\equiv&\sqrt{-c}\,q  \nonumber \\
 &p\equiv&-a/c^{2}       \nonumber \\
 &a_{n}\equiv&c\sqrt{p}\,A_{n}\;\;\;\;n=0,1,2\ldots
\label{eqnarray:paramredf}
\end{eqnarray}
 Note that $a<-c^{2}/8$ corresponds to $p>1/8$. Then eq.(\ref{eq:feper3})
 reads
\begin{equation}
 f\left(q,A_{0},A_{1},\ldots\right)=pc^{4}\left[\sum_{n=0}^{\infty}\left(
 q^{4}n^{4}-q^{2}n^{2}-p\right)A_{n}^{2}+\frac{p}{2}\left<v^{4}\right>\right].
\label{eq:feper4}
\end{equation}
 Here the function $v=v(z)$ is one of the Fourier expansions
 (\ref{eq:fsI})-(\ref{eq:fsIII}) with the amplitudes $a_{0},a_{1},.\ldots$
 replaced by the amplitudes $A_{0},A_{1},.\ldots$ etc. The same redefinition
 of $all$ amplitudes is applied to (\ref{eq:fsIV}).
 Note that $p$ is the only control parameter which enters into the free energy.

 In order to find out the configurations which minimize the free energy
 functional it remains to minimize the expression (\ref{eq:feper4}) with
 respect to the amplitudes $A_{1},A_{2},...$.
 The details of numerical procedure for achieving this
 aim are as follows \cite{kn:Dananic}.
 Since there are no periodic solutions of equation (\ref{eq:EL}) having
 a finite number of harmonics in the Fourier expansion, one has to
 choose a finite number ($N$) of first harmonics
 as a starting approximation. Unfortunately there is no
 systematic way to choose appropriate initial approximations, so that we
 usually start by reasonable guesses, like those mentioned in the previous
Section.
 Then one varies over these harmonics until the functional acquires a local
 minimum in the $N$-dimensional space $A_{1},A_{2},...,A_{N}$. The resulting
 configuration is the corresponding Fourier expansion $v_{N}(x)$. If the last
 coefficient in this expansion, $A_{N}$, is not small enough in comparison
 with the first leading harmonics $A_{1},A_{2}$...,
 one enlarges the space of coefficients
 by further $M$ harmonics. Taking again as the initial set first $N$ harmonics
 defining $v_{N}(x)$ and additional $M$ harmonics chosen by some guess, one
 repeats the variational procedure until the criterion for the small enough
 value of the last coefficient in the Fourier series is satisfied. Our usual
 criterion is that the last coefficient should by at least $10^{-13}$ times
 smaller than the values of leading coefficients. Usual starting number of
 harmonics is about ten. In order to get periodic solutions which will be
 shown in the next Section
 we need at least fifty harmonics. Besides
 the periodic configuration $v(x)$ and the corresponding value of the free
 energy, we also obtain the wave number $q_{0}$ given by eq.(\ref{eq:mink}).
 Thus we have the complete characterization of a given periodic solution.

 In the second step we investigate how well the obtained configuration
 satisfies the EL equation (\ref{eq:EL}). Firstly we insert a given periodic
solution
 into eq. (\ref{eq:EL}) and look how its left-hand side varies with $z$. It
 fluctuates around zero, with fluctuations which are chaotic and of
 characteristic magnitude as small as $10^{-7}$. Thus the periodic solutions
 obtained through the variational calculation on the restricted space of
 periodic functions indeed well satisfy the EL equation. The same
 conclusion follows from  the second check in which we take a periodic
 solution as an initial approximation for the Newton-Kantorovich iteration.
 For this purpose we convert the Fourier
 expansion into a chosen basis of spline functions. The Newton-Kantorovich
 procedure ended, as a rule, after the first iteration.

 The third numerical task concerns the local stability of obtained solutions.
 It is analyzed by diagonalizing the Hess determinant in the space of
 Fourier coefficients $A_{n}$. Instead of being eliminated by making use
 of eq.(\ref{eq:mink}), the wave number is treated as an additional
 variable, since then the determinant can be much easier calculated.
 As far as the lowest eigenvalue of
 this determinant is positive the configuration is considered to be stable.
 However, this stability concerns only periodic variations. The stability under
 general variations could, in principle, be studied by making use of the Bloch
 theorem for the equation (\ref{eq:eig}). In such cases, one
 would have to diagonalize
 a matrix of two times larger order than for periodic variations. In addition,
 the eigenvalues of this matrix depend on additional, Bloch's, wave vector.
 We have not performed such
 diagonalizations, $i.e.$ the boundaries of stability given
 in the next section are calculated
 only with respect to the periodic variations.
\newpage
\section{The structure of the phase diagram}
\label{sec-phasediagram}
 It was already pointed out in the previous Section that $p=-a/c^{2}$ is
 the only combination
 of original physical parameters which remains after convenient redefinition of
 scales in the free energy functional and  the EL equation.
 Since the variations of parameters $a$ and $c$ are the most relevant for the
 IC transition, this means that all critical lines in the phase diagram
 represented by the ($a,c$) plane are parabolas determined by critical values
 of $p$. All parabolas meet with a common tangent at the origin (0,0), named
 a "multicritical Lifshitz point"
 \cite{kn:Michelson,kn:Aharony}.

 Further insight into the phase diagram can be gained from the dependence
 of the free energy on the parameter $p$ for particular configurations. Let
 us rewrite the expression (\ref{eq:feper4})  in the form
\begin{equation}

f[v]=\frac{|a|c^{2}}{bd}\left<q^{4}v''^{2}-q^{2}v'^{2}-pv^{2}+\frac{|p|}{2}v^{4}\right>,
\label{eq:feper5}
\end{equation}
 choose
\begin{equation}
 f_{0}=\frac{|a|c^{2}}{bd}
\label{eq:feunit}
\end{equation}
 as the energy unit, and write $f/f_{0}\rightarrow f$ further on.
 The free energy for the simplest homogeneous solutions
\begin{equation}
 v=\pm1
\label{eq:hom}
\end{equation}
 then reads
\begin{equation}
 f[\pm1]=\frac{-p}{2}.
\label{eq:fehom}
\end{equation}
 The approximate sinusoidal solution
 which figures in the phase diagrams of
 Refs.\cite{kn:Michelson,kn:Aharony} is given by
\begin{equation}
 v(z)\simeq\sqrt{2}\sqrt{\frac{2}{3p}\left(p+\frac{1}{4}\right)}\,\cos(z),
\label{eq:perapp}
\end{equation}
 with the corresponding energy
\begin{equation}

\frac{f\left[v(z)\right]}{f_{0}}=-\frac{1}{3p}{\left(p+\frac{1}{4}\right)}^{2},
\label{eq:feperapp}
\end{equation}
 and the "scaled" wave number
\begin{equation}
 q^{2}\simeq\frac{1}{2}
\label{eq:wavevecperapp}
\end{equation}
 (the actual wave number is given by equation (\ref{eqnarray:paramredf})).
 The more precise result for this configuration obtained through the
 minimization method of previous Section is presented in Table Ia for $p=1$.
 The solution $u(z)$ and its ($u,u'$) section are shown in Fig.1a,b. As it was
 already stated in Ref.\cite{kn:Michelson}, the ratios $|a_{3}/a_{1}|$,
 $|a_{5}/a_{1}|$ etc. are rather small, while the values of $a_{1}$ and
 the free energy $f$ are close to those given by eqs. (\ref{eq:perapp})
 and (\ref{eq:feperapp}) ($i.e.$ 0.9128 $vs$ 0.913 and -0.523 $vs$ -0.5208
 respectively). Furthermore the numerical value of the parameter $p$ for
 which the free energy becomes equal to the free energy of the homogeneous
 phase (\ref{eq:hom}), $i.e.$ for which the phase transition of the first
 order occurs, is
\begin{equation}
 p_{T}\simeq 1.177,
\label{eq:expICT}
\end{equation}
 again close to the approximate value 1.122. The wave vector corresponding
 to $p_{T}$ is only slightly smaller than that given by
 eq.(\ref{eq:wavevecperapp}). Furthermore the solution of Fig.1a is
 thermodynamically stable for
\begin{equation}
 p<p_{C}=1.835
\label{eq:psinmetast}
\end{equation}
 which is close to the estimaton $p_{C}=2$ given in \cite{kn:Michelson}.

 The homogeneous (\ref{eq:hom}) and (almost) sinusoidal (eq.(\ref{eq:perapp})
 and Table I) configurations represent the sceleton of the phase diagram
 in the range $a<0,c<0$. Other, more complex configurations bring a fine
 structure into this diagram. We start with a configuration
 belonging to the class ($ii$), presented in Table Ib and Fig.1b.
 The appearance of a small circle in Fig.1b indicates that this configuration
 emerges through a local mixing of the incommensurate sinusoidal configuration
 from Fig.1a and the commensurate homogeneous configuration $v=1$.
 Its free energy lies between the energies of these two
 configurations. The thermodynamic stability of this configuration is limited
 to a rather narrow range $p_{L}<p<_{R}$ with $p_{L}=0.956$ and $p_{R}=1.05$.
 This is to be contrasted to the homogeneous configuration
 which is stable for every $p\geq 1/8$ and the sinusoidal configuration which
 is stable in the range of $-1/4\leq p\leq 1.835$.
 The wave number of the new configuration varies from $q(p_{L})=0.42$
 to $q(p_{R})=0.4$, $i.e.$ it is considerably smaller than the wave number
 of sinusoidal configuration (eq.(\ref{eq:wavevecperapp}))
 everywhere in the domain of its stability. We
 also note that the right edge of stability is only slightly above the crossing
 of this line with that of the homogeneous ferromagnetic
 configuration (see also Fig.2).

 Next in the hierarchy of complexity is the configuration which
 again falls into the class ($i$).
 It has two small circles in section plane $(v,v^{\prime})$, one for each
 fixed point $v=\pm 1$. Since the mixing of the homogeneous and sinusoidal
 solution is even stronger in this case, the corresponding energy is higher
than
 that of configuration from Fig.1b. The representative of this
 configuration in the ($x,v$) and ($v,v^{\prime}$) planes is
 shown in Fig.1c. Again, the range of
 its stability is very narrow, and approximately of the
 same width as for the preceding configuration ($p_{L}=0.8, p_{R}=0.98$).
 The number of coefficients needed to accomplish the criterion established
 in Sec.3 for this and further configurations is larger than fifty. We do not
 present them for the sake of space.

 The further two configurations falling
 into the class ($i$) and ($ii$) are shown in Figs.1d,e respectively. The
simplest
 configuration from the class ($iii$) is shown in Fig.1f.
 Finally, in Fig.1g we
 present one solution from the class ($iv$). This solution has a lower symmetry
 than others. Namely, there is no point on $x$ axis with respect to which
 it is either even or odd, opposite to other solutions for which at least
 one such point exists. The configuration $u(x)$ belonging to class ($iv$) is
 different (in the sense given above) from $u(-x)$, but the free energies
 and the periods of both $u(x)$ and $u(-x)$ are same.

 All presented (Figs.1a-g), and hopefully further more complex, configurations
 could be viewed as being built from the following basic blocks:
 commensurate domains (left and right small circles in the section plane
 ($v,v'$)), and the half-periods of sinusoidal configuration (left and
 right halves of large ellipses in the section plane ($v,v'$)). Both types of
 blocks have, up to very slight deviations, same lengths in all derived
configurations.
 Thus, we come to some kind of a nonlinear superposition principle.
 In that respect one can introduce a systematization of all periodic
 configurations, by using following scheme: Let us denote the blocks from
 Figs.2a,b,c,d by letters $s_{+},s_{-},d_{+}$ and $d_{-}$ respectively. Any
 periodic configuration is then designated by a $word$ in which the order
 of letters respects a simple rule by which a letter with +(-) subscript
 can be followed only by a letter with -(+) subscript. Taking this rule
 into account, one can further simplify the lettering by omitting subscripts.
 E.g. the sinusoidal configuration from Fig.1a is denoted by
 ...$s_{+}s_{-}s_{+}s_{-}$..., or in a condensed form $s_{+}s_{-}=s^{2}$,
 where the last word corresponds to one period of the configuration. Note
 that one condensed word must have an even number of letters. There are three
 different configurations with two-letter words, namely $s_{+}s_{-}=s^{2}$,
 $s_{+}d_{-}=sd$ and $d_{+}d_{-}=d^{2}$. They are presented in Figs.1a,b,c
 respectively. The "four-letter" configurations are
$s_{+}s_{-}s_{+}d_{-}=s^{3}d$,
 $s_{+}s_{-}d_{+}d_{-}=s^{2}d^{2}$ and $s_{+}d_{-}d_{+}d_{-}=sd^{3}$. We
 found two of this three configurations, $i.e.$ the first (Fig.1d) and
 the second (Fig.1e). Finally, the word for the configuration from
 Fig.1g has twelwe letters and reads
$s_{-}s_{+}s_{-}s_{+}s_{-}s_{+}s_{-}s_{+}d_{-}d_{+}s_{-}d_{+}=s^{9}d^{3}$.

 Unfortunately, at the time being we do not have any efficient numerical
 method which would enable the determination of a configuration which
 corresponds to any chosen word.
 Still, one could imagine a word with arbitrary many letters, $i.e.$ a
 sequence composed of sinusoidal and commensurate domains ordered in a
 random way inside one long period. In other words, the great freedom in the
 formation of macroscopic domain patterns may be in principle still
 interpreted in terms of "determinism" presented in the EL equation
 (\ref{eq:EL}). However, such patterns are obviously unreachable numerically.
In this
 respect, we should also mention that we did not find any configuration
 in which one commensurate domain is followed by another without inserting
 at least one half-sinusoidal block, $i.e.$ a configuration which would
 have in the ($u,u'$) section two small circles in succesion.
 Furthermore, the commensurate domains with larger lengths
 could be expected for the parameters $p$ larger than the value given by
 eq.(\ref{eq:expICT}). However, it comes out that by adding more and more
 harmonics the wave number of such configurations does not stabilize
 to a finite value but continuously goes to zero.

 To summarize this Section, we show in Fig.3 the dependence of the free
 energy on the parameter $p$ for all presented solutions. The corresponding
 limits of stability are gathered in Table II. Two lines which dominate
 in Fig.3 belong to the commensurate and sinusoidal configurations. The
 relevant range of values for $p$ is limited from the left side by the edge
 of thermodynamic instability of commensurate configuration ($p>1/8$) and
 from the right side by the critical value for the transition from the
 commensurate to the sinusoidal configuration ($p<1.177$). Inside this
 range there are (probably infinitely) many relatively short lines
 belonging to periodic configurations composed of blocks from Fig.8. As a
 rule, all these short lines cross the commensurate line on their right ends
 and cease to be thermodynamically stable immediately after crossing it.
 The periods of the configurations from  Table II are shown in Fig.4. This
 diagram suggests that the separate curves are just parts of a smaller number
 (perhaps two)
 curves representing multivalued dependences of the period on the parameter
 $p$. (Note that the periods of configurations with longer words would be
 situated higher in the diagram). The missing parts on these curves
 would belong to unstable periodic configuration. Note that such
 configurations are out of scope of our numerical algorithm.

\newpage
\section{Conclusions}
\label{sec-sum}
 The discommensurations, $i.e.$ the objects connecting commensurate domains
 with different or equal phases ($n\geq 3$) or signs ($n=1,2$), have a
 central role in the uniaxial IC transitions. The present analysis shows
 that in the systems of the II class they enter into the phase diagram as
 ingredients of metastable configurations illustrated by Figs.1b-g. As was
 already pointed out in the previous Section, the discommensurations in
 these configurations just coincide with the local incommensurate order.
 Numerical results indicate that as a rule such configurations may be
 locally stable only within the range of coexistence of the homogeneous and
 the incommensurate (almost sinusoidal) ordered states. This is in contrast
 to the class I in which the sinusoidal incommensurate and the commensurate
 configurations do not coexist, but instead pass gradually from one to
 another through the continuous family of soliton lattices, with the phase
 solitons playing the role of discommensurations. By approaching the IC
 transition the distances between two phase solitons ($i.e.$ the lengths
 of commensurate domains) tend towards infinity, so that a single
 discommensuration is well defined. It is also unique, since there is only
 one type of separatrix which join two hyperbolic fixed points which
 correspond to homogeneous (commensurate) state(s) in the sine-Gordon
 problem. Our numerical results suggest that this is not a case for the
 II class, $i.e.$ the isolated discommensurations do not appear as stable
 objects, since all periodic domain configurations cease to be stable at
 finite periodicities. Evenmore an isolated discommensuration is not
 uniquely defined, since there are many ways to join two fixed points
 ($u=\pm1,u'=u''=u'''=0$) in Figs.1b-g due to the complex Lyapunov
 exponents characterizing their stability.

 The presence of metastable configurations makes the IC transitions for the
 II class more complex than it was thought before [3,13-15].
 In fact, being the "droplet" states which mix two basic states of the
 1st order transition, these configurations complete in a natural way the
 phase diagram from Fig.3. It should be however stressed that they are
 purely intrinsic solutions of EL equation for the free energy functional
 (\ref{eq:feren}). In other words, there is no need for some extrinsic inputs
 (like boundaries, external field(s), defects, impurities, etc.), which are
 usually necessary for the stabilization of mixed states in the systems
 passing through the phase transition of the 1st order. This however does
 not exclude a possible additional extrinsic influences onto the
 "microscopic" periodic configurations, which, as will be argued below,
 very probably take place in real systems.

 Let us now consider possible physical implications of new metastable lines
 in the phase diagram from Fig.3. At first, a direct identification of a
 periodic domain configuration in structural measurements implies a detailed
 knowledge of corresponding structure factor. Our preliminary numerical
 computations suggest that the structure factors of domain configurations
 from Figs.1b-g and that of the sinusoidal configuration (Fig.1a) are hardly
 distinguishable in the experimentally most interesting Brillouin zones
 with rather low Bragg indices. More noticeable differences are however
 expected in the zones with large indices in which the scattering from the
 domain configuration is on the fine scale more noisy than that from the
 sinusoidal configuration. Thus, the structural identification of complex
 domain patterns could be a subtle experimental task. Still, if the
 commensurate and/or incommensurate segments are rather long, the
 corresponding structure factor has coexisting peaks at $q=0$ and
 $q=\pm\sqrt{-c/2d}$. Such coexistence of peaks was indeed observed in the
 neutron scaterring measurements \cite{kn:Aime} on the diacetylene
 bis-p-toluene sulphonate of 2.4-hexadyne-1.6 diol (PTS), the system which
 belongs to the class II and passes through the Lifshitz point by
 gradual conversion from the monomer to the polymer crystal structure
 \cite{kn:Patillon,kn:Pouget}.

 The physical properties of a given system are usually followed by varying
 one or two parameters, e.g. temperature and another appropriate
 quantity like pressure, strain, concentration of some constituent, degree
 of polymerization, etc. Any such variation can be in principle represented
 by a path in the ($a/b$,$c/d$) plane, $i.e.$ by some variation of the
 reduced control parameter $p$. Depending on the details of such a path
 (including its direction), and on other possible specific influences,
 the system may pass through a number of metastable periodic configurations,
 showing so discontinuities in physical properties like the staircase
 dependence on e.g. temperature, the hysteresis behavior, the dependence
 on the initial state ($i.e.$ history), etc. Such effects are indeed often
 met in the systems of the II class \cite{kn:Cummins}. The well-known examples
are thiourea
 \cite{kn:Denoyer}, silicium dioxide (quartz) \cite{kn:Dolino} and again
 PTS \cite{kn:Aime,kn:Patillon,kn:Pouget}. The hysteresis and memory effects
 in thiourea \cite{kn:Jamet} and some other ferroelectric materials like
 barium sodium niobate (BSN) \cite{kn:Toledano1} were most often interpreted
 by assuming that impurities are mobile enough to form a density wave as a
 response to the incommensurate lattice modulation, which in reverse tends
 to pin the modulation wave and to fix its position in the crystal
 \cite{kn:Lederer,kn:Errandonea}. On the contrary, the explanation of these
 effects based on the present analysis is essentially microscopic
 and universal for the whole II class of IC systems \cite{kn:zamjedba}.
 The defects (and/or other possible extrinsic causes) may still have a
 secondary role as triggers which favor the stabilization of some particular
 periodic domain patterns among all those available from the phase diagram
 from Fig.3. Our explanation however does not invoke their mobility. Note
 that the dominant defects in some systems exhibiting a global hysteresis
 are by their nature imobile, like e.g. those in randomly polymerized
 crystals of PTS. The polymerization in PTS acts simultaneously as a control
 parameter (through the variation of the parameter $c$) and as a
 stabilization mechanism for domain patterns \cite{kn:Pouget}. On the other
 hand the present model cannot be directly applied without further extensions
 on quartz, although this material exhibits very dinstinct discontinuities in
the wave
 number and birefrigence \cite{kn:Dolino,kn:Dolino1}, including an
 intriguing time dependence of staircase steps \cite{kn:Mogeon}. At first,
 the incommensurate modulation in quartz is not unimodular but
 characterized by a star of six equally intense satellites \cite{kn:Dolino2}.
 Secondly, it is hard to say anything about the kinetics of incommensurate
 states without a more complete insight into the topological barriers
 between various periodic domain patterns.

 In conclusion we stress a theoretical significance of the considered Landau
model.
 When extended with transverse gradient terms, it represents the Landau
 expansion for the e.g. ANNNI model in the vicinity of the Lifshitz
 point \cite{kn:Selke}. The phase diagram of Fig.3 gives a new insight
 into this range of ANNNI phase diagram. Note that periodic domain patterns
 from Fig.3
 are not directly linked to the stable commensurate modulated configurations
 which appear far enough from the Lifshitz point in the ANNNI model
\cite{kn:Selke}.
 Mathematical properties of the model considered here are particularly
 challenging. Besides being very probably nonintegrable, it has an
 unconventional classical mechanical counterpart with the kinetic energy
 which is not positively definite (in contrast to the class I models which
 do have such a counterpart because of the absence of
 higher derivative terms of the order parameters). As a consequence, the
 periodic solutions which are especially important due to their
 thermodynamic stability, have a particular type of orbital instability.
 They seem to be immersed into a "chaotic web" which (again very probably
 due to the indefinitiveness of the kinetic part of the Hamiltonian), is
 even not localized in the phase space. In that respect one encounters an
 intriguing open question of critical fluctuations in a purely (or quasi)
 one-dimensional version of the model, for which the renormalization
 group expansion \cite{kn:Hornreich,kn:Aharony} cannot be applied.
 This and other already mentioned extensions of the model are possible
 subjects of future investigations.

 \underline {Acknowledgements}

 We are indebted to E. Coffou and M. Rogina for the help in the numerical
 analyses. A. B. acknowledges interesting conversations with D. K. Campbell
 and J. P. Pouget.

\newpage
\begin{center}
 Figure Caption
\end{center}

Fig. 1\,\, $u(z)$ dependence (left) and ($u,u'$) section (right)
 of the configurations $s^{2}$ (a), $sd$ (b), $d^{2}$ (c), $sd^{3}$ (d),
 $s^{2}d^{2}$ (e), $s^{2}d^{4}$ (f) and $s^{9}d^{3}$ (g). For the configuration
 (c) $p=0.9$, and for all other configurations $p=1$. The symbolic words for
 configurations are defined in the text.

Fig. 2\,\, The building blocks $s_{+}$ (a), $s_{-}$ (b), $d_{+}$ (c) and
 $d_{-}$ (d) for configurations from Fig. 1.

Fig. 3\,\, The free energy  $f$ {\underline vs} the control parameter $p$ for
the
 configurations from Figs (1a-f) and the homogeneous configuration (dashed
line)

Fig. 4\,\, The periods of configuration from Figs (1a-d) {\underline vs} the
control
 parameter $p$.

\vspace{2cm}
\newpage

\begin{center}
 Table Captions
\end{center}

Table I\,\, The Fourier coefficients for the configurations $s^{2}$
 (Ia) and $sd$ (Ib).

Table II\,\, The values of parameter $p$ for the left and right end
 points of stability ranges for the configurations from Figs 1a-f.

\newpage

\begin{center}
 {\bf Table I }
\vspace{5 mm}

\begin{tabular}{|c|c|c|c|}
 \hline
 \multicolumn{4}{c} {\rm (Ia)} $s^{2}$ \\ \hline
 $n$   & $A_{n}$ & $n$ & $A_{n}$ \\  \hline
  1    &$0.925$ & 9 & $1.969\cdot 10^{-8}$ \\  \hline
  3    &$-2.551\cdot 10^{-2}$ & 11 & $-1.568\cdot 10^{-10}$ \\  \hline
  5    &$2.444\cdot 10^{-4}$  & 13 & $1.210\cdot 10^{-12}$ \\  \hline
  7    &$-2.314\cdot 10^{-6}$ & 15 & $-9.125\cdot 10^{-15}$ \\  \hline
 \multicolumn{4}{c} {\rm (Ib)} $sd$ \\ \hline
 $n$   &$A_{n}$ & $n$ & $A_{n}$  \\  \hline
  0    &$0.351$ & 9 & $9.733\cdot 10^{-6}$ \\  \hline
  1    &$0.794$ & 10 & $4.385\cdot 10^{-6}$ \\  \hline
  2    &$-0.381$ & 11 & $-1.539\cdot 10^{-6}$ \\  \hline
  3    &$7.873\cdot 10^{-3}$ & 12 & $1.454\cdot 10^{-7}$ \\  \hline
  4    &$3.323\cdot 10^{-2}$ & 13 & $4.136\cdot 10^{-8}$ \\ \hline
  5    &$-9.883\cdot 10^{-3}$ & 14 & $-1.668\cdot 10^{-8}$ \\  \hline
  6    &$6.204\cdot 10^{-4}$ & 15 & $1.899\cdot 10^{-9}$ \\  \hline
  7    &$3.960\cdot 10^{-4}$ & 16 & $3.489\cdot 10^{-10}$ \\  \hline
  8    &$-1.288\cdot 10^{-4}$ & 17 & $-1.689\cdot 10^{-10}$ \\  \hline
\end{tabular}
\end{center}

\vspace{5 mm}
\begin{center}
 {\bf Table II}
\vspace{5 mm}

\begin{tabular}{|l|l|r|}                  \hline
 $word$         &\,\,\,$p_{L}$   &\,\,\,$p_{R}$  \\  \hline
  $s^{2}$       &$-0.25$         &$1.835$    \\      \hline
  $sd$          &$\,0.95661$     &$1.082$    \\      \hline
  $d^{2}$       &$\,0.815$       &$0.989$    \\      \hline
  $sd^{3}$      &$\,0.894$       &$1.020$    \\      \hline
  $s^{2}d^{2}$  &$\,0.970$       &$1.080$    \\      \hline
  $s^{2}d^{4}$  &$\,0.900$       &$1.050$    \\      \hline
\end{tabular}
\end{center}

\end{sf}

\begin{thebibliography}{99}
\bibitem{kn:surface} The similar procedure can be applied also when only the
first derivative of the
 order parameter is involved $i.e.$ in the Landau expansion appropriate for
systems of the class I.
 In that case the simplest choice of "surface" contribution would suggest the
incommensurate plane
 wave configuration described by a complex order parameter. There, the
oscillations would affect only
 a phase of the order parameter since the Lifshitz invariant does not contain
the derivative of its amplitude.
\bibitem{kn:zamjedba} The global hysteresis in systems which do not belong
                      to the II class can be also interpreted as a microscopic
                      phenomenom already in the weak coupling limit, after
                      including into the sine-Gordon model at least two
                      dominating commensurate potentials with different
                      periods. ( A. Bjeli\v{s}\ and M. Latkovi\'{c} , to be
                      published)
\end{thebibliography}
\end{document}